\documentclass[preprint]{vgtc}
\usepackage[T1]{fontenc}
\usepackage{times}
\usepackage{amsmath}
\usepackage{booktabs}
\usepackage{url}
\usepackage[hidelinks,bookmarks]{hyperref}

\usepackage{array}
\usepackage{tabularx}
\usepackage{longtable}
\patchcmd{\thebibliography}{\scriptsize}{\small}{}{}
\preprinttext{}
\author{Swethasree Sribabu\textsuperscript{1}\thanks{Corresponding author. E-mail: \href{mailto:am24s015@smail.iitm.ac.in}{am24s015@smail.iitm.ac.in}}
\and Vishnu Raj\textsuperscript{2}
\and Gouthaman KV\textsuperscript{2}
\and Manivannan Muniyandi\textsuperscript{1}}
\shortauthor{Swethasree Sribabu, Vishnu Raj, Gouthaman KV, Manivannan Muniyandi}
\affiliation{\scriptsize \textsuperscript{1}Indian Institute of Technology Madras\\\textsuperscript{2}Dolby Laboratories}

\title{FRAME VR: Facial-Reaction-Adaptive Movie Engine in VR}
\AtBeginDocument{\hypersetup{pdftitle={FRAME VR: Facial-Reaction-Adaptive Movie Engine in VR},pdfauthor={Swethasree Sribabu, Vishnu Raj, Gouthaman KV, Manivannan Muniyandi},pdfsubject={},pdfkeywords={}}}
\abstract{A face-tracking headset already measures facial movement while a person watches, so a film could respond to its viewer without asking. We present FRAME VR, a face-responsive film system that runs entirely on a standalone virtual reality headset. At each cut, a per-viewer rule compares the viewer's nose wrinkling and upper-lip raising with their own resting face and chooses the next scene from those not yet shown. A compact expression model runs concurrently with playback and logs its predictions. With 36 viewers and three films, the next scene moved in the requested direction, as judged by pilot disgust ratings, on 36 of 51 choices with a usable reading (71\%), or 36 of 108 choices overall (33\%). Of the remaining choices, 57 followed the preset order because no reading was usable, and 15 had a reading but did not meet the request. Most viewers did not report noticing a response. Only 4 of the 35 respondents thought any film had responded to them, while 25 were unsure and 6 said no. We also present a scene check that filmmakers can run before any viewer sees the film. It shows how one responsive cut can use up the time needed to read the next reaction or the last gentler scene. After adjusting for film and presentation position, the comparison of whether the intensity suited viewers was inconclusive. We draw five lessons for filmmakers: give the story a stable opening and ending, leave time to read the viewer after each cut, save a scene for the next request, check what played and ask what viewers understood, and ask what viewers want as well as what they feel.}

\begin{document}
\maketitle
\section{Introduction}
\label{sec:intro}

Two people can watch a restaurant inspector open the same container of old food. One may want the film to move on to something gentler, while the other may find the scene barely affecting. A film could use their facial movements to choose what to show next without interrupting the story to ask. A face-tracking virtual reality (VR) headset already estimates movements such as a nose wrinkle or an upper-lip raise. These estimates were designed to animate an avatar, but they are also available while someone watches a film. We made the choice concrete with one kind of content: disgust at spoiled food and poor hygiene. How much a person's face moves, how much disgust they feel, and how much they want to feel are separate questions.

\begin{figure*}[t]
\centering
\includegraphics[width=0.80\textwidth]{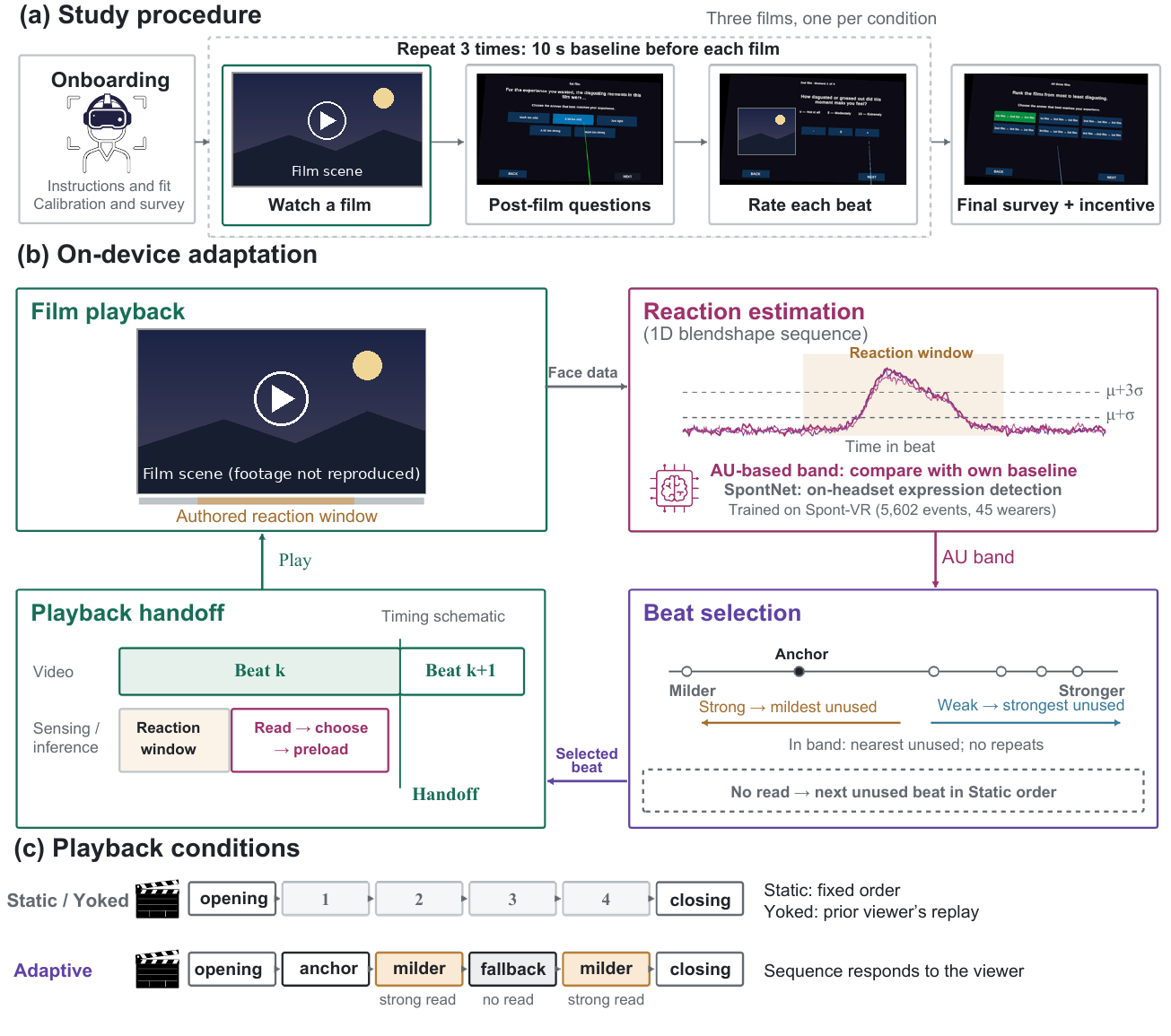}
\caption{How FRAME VR runs on a standalone headset. (a) Each viewer watches three films, rates each film and its individual scenes, and answers a final survey. A 10 s resting-face recording precedes each film. (b) A filmmaker marks when to read the face within each scene. The diagram calls a scene a beat and the fixed first scene the anchor. The rule compares the peak of the nose wrinkle and upper-lip raise (facial action units AU9 and AU10) with the viewer's resting-face mean $\mu$ and standard deviation $\sigma$. The result, called the AU band in the diagram, is a weak, middle or strong reading. The per-viewer rule then chooses from unused scenes while the current scene ends. Fallback marks a preset choice made when there is no usable reading. SpontNet only logs expression predictions and does not choose scenes. The curve, the timing and the screen image are schematic. (c) Static (fixed order) plays a fixed order. Adaptive (face-responsive) uses the current viewer's reading or the next unused scene in fixed order when there is no reading. Yoked (stored-order replay) replays a stored order for the same film, or one of three preset orders when no earlier viewer's order was available.}
\label{fig:system}
\end{figure*}

We built FRAME VR, a face-responsive film system that runs entirely on a standalone headset (Figure~\ref{fig:system}). The film starts and ends the same way for every viewer, but the scenes in the middle can change order. The filmmaker marks the part of each scene where the headset should read the viewer's face. When the headset gets a usable reading, a per-viewer rule compares the movement with the same viewer's resting face. We chose this simple rule on purpose, because it compares each viewer only with themselves and lets us trace every choice to one reading. A strong reading asks for a gentler scene, a weak reading asks for a stronger one, and a reading in between asks for a scene of similar intensity. Alongside playback, the headset also runs SpontNet, a compact expression model with 105,617 parameters trained on 5,602 events from 45 headset wearers. It logs its predictions and does not choose scenes.

In a study with 36 viewers, the rule delivered the requested direction on 36 of 51 choices with a usable reading (71\%), or 36 of 108 choices overall (33\%) in the face-responsive condition, which we call Adaptive. Here, a delivered direction means that the next scene's pilot rating moved the way the reading asked. Without a usable reading, 57 choices followed the preset order, and another 15 had a reading but did not meet the request. The order could change even without a reading, because the fallback took the next unused scene, and sometimes that scene was stronger. After adjusting for film and presentation position, whether the intensity suited viewers did not differ reliably between conditions. Delivering a change also did not mean that viewers noticed it, and only 4 of 35 respondents thought any film had responded to them. This gap suggests that a film can adapt without viewers noticing, so what played and what viewers report need separate checks.

Better face tracking would solve only part of the problem, because the scene lists also limit the rule. In two of the three films, the gentlest option after the fixed first scene leaves too little time to read the next reaction. In the third film, the same choice leaves a long enough interval to attempt a reading but uses up the last gentler scene. We turned these examples into a scene check that a filmmaker can run on the scene list and reading intervals before showing the film to anyone.

Our contributions are:
\begin{itemize}
\item \textbf{A face-responsive film system on a standalone headset.} FRAME VR reads the face, chooses among unused scenes, and plays the film on the headset alone. For each cut it records whether the reading was usable, which direction was requested, and which scene played, so delivery can be followed from request to screen.
\item \textbf{Delivery and perception findings.} The rule delivered most requests with a usable reading (36 of 51), though only 36 of 108 choices overall, and most viewers did not report noticing a response. We compare what viewers reported with what played.
\item \textbf{A scene check.} After the study, we wrote a check that follows every point in each film, every kind of reading and every tie in the existing scene lists. It separates a scene that suits the current request from one that also leaves time for the next reading and a scene to act on it.
\item \textbf{Lessons for filmmakers.} At each cut, filmmakers need to check the requested direction, the time left to read the viewer and the scenes that remain, all at once. We draw five lessons from the delivery record, the edits and the viewer reports.
\end{itemize}

\section{Related work}
\label{sec:related}

\subsection{Films that respond without asking for a choice}

Viewers have helped shape films through physiological signals, brain activity, gaze, and facial movement. Tikka's enactive cinema and Kirke and colleagues' cinema deployment use the viewer's body to help determine what plays~\cite{tikka2008enactive,tikka2012enactive,kirke2018unconsciously}. \emph{The MOMENT} uses brain activity to edit a film as it plays~\cite{ramchurn2019director}. \emph{Before We Disappear} is the closest example to our application. It selects scenes using camera-based estimates of a viewer's facial response relative to a baseline~\cite{ramchurn2024before}. Cin\'evoqu\'e uses gaze and time to change an immersive live-action narrative~\cite{pillai2019cinevoque}. VR experiences have also used a deliberate smile or facial muscle activity as input~\cite{li2025forced,reidy2020facial}. FRAME VR instead uses the facial movements a headset already tracks while people watch a film on a virtual screen.

Tools for writing interactive stories check which story states and interactions are compatible~\cite{mateas2003facade,riedl2013,kapadia2015authoring}. Finite question pools in adaptive testing provide another example of choices becoming unavailable as items are used~\cite{vanderlinden1998model,huitzing2005infeasibility}. Applied to the cut between scenes, both ideas ask whether an unused scene can move in the requested direction, and whether the choice leaves time to read the viewer's next reaction and a scene to act on it. We answer this from the scene lists and delivery records, without relying on viewer ratings.

\subsection{What a face and a viewer can each tell us}

Headsets estimate facial movements through coefficients originally intended to animate an avatar~\cite{bai2024universal,cannavo2019animation,zhang2023facial}. EmoHeVRDB records people deliberately performing expressions in a headset~\cite{ortmann2024emojiherovr}. Comparisons of posed and spontaneous reactions show that the setting matters~\cite{tare2025detecting,rahman2026generalized}. Work on image-based micro-expressions offers ways to find brief movements, and warns that test and training data can leak into each other~\cite{yan2014casme,davison2018samm,li2023cas,varanka2024data}. BlendME uses headset recordings and an avatar replay that the viewer can annotate after watching~\cite{sribabu2026blendme}. Our model, SpontNet, was trained in a separate headset recording study and only logs predictions in the film system. None of these works uses spontaneous headset movement to choose film scenes.

Film is an established way to elicit disgust~\cite{gross1995emotion,schaefer2010assessing,samson2016eliciting}. Nose wrinkling and upper-lip raising can accompany disgust~\cite{kunz2013pain,vrana1993psychophysiology}, and reported feelings and facial movements can diverge~\cite{duran2021emotions,reisenzein2013coherence,barrett2019reconsidered}. Viewers may also seek out an unpleasant sensation as part of a film~\cite{andersen2020playing,kiss2024role,rozin2013glad,benford2012uncomfortable}. We asked separately what they felt and whether the intensity suited them.

Knowing or believing that a system responds can itself shape a person's experience~\cite{denisova2015placebo,peng2018trip,sehrt2024closing}. A stored replay lets the same film vary without taking input from the current viewer. This comparison has a familiar constraint: the recording carries the earlier viewer's path and circumstances with it~\cite{church1964systematic}. We checked that each stored-order replay played its reserved order, the order fixed for that session before it began, and asked viewers separately whether they thought a film responded to them.

\begin{figure*}[t]
\centering
\setlength{\tabcolsep}{7pt}
\renewcommand{\arraystretch}{1.22}
\begin{tabularx}{0.94\textwidth}{@{}>{\raggedright\arraybackslash\bfseries}p{0.17\textwidth}>{\raggedright\arraybackslash}X>{\raggedright\arraybackslash}X@{}}
\toprule
Part of the study & What we built or collected & What we checked \\
\midrule
Film editing & 3 films with fixed openings and endings, pilot-rated scenes, and times marked for reading the face & Whether the scene list can supply a change and leave time and material for the next choice \\
Headset playback & An application that reads the face, picks an unused scene, and starts loading it before the cut & Usable readings, requested directions, scenes played, and timing of successful model runs \\
Film deployment & 36 viewers who each watched a fixed-order film, a face-responsive film, and a stored-order replay & Scene delivery, experience ratings, and whether viewers thought a film responded to them \\
Separate expression study & 5,602 annotated movements from 45 headset wearers, used to train SpontNet & Recognition on wearers held out during training, with predictions logged alongside the film rule \\
\bottomrule
\end{tabularx}
\caption{What was built and evaluated. Read across each row for what was built and how it was checked. The film deployment and the facial-expression study used different participants. Only the per-viewer rule chose scenes, using nose and upper-lip movements, and SpontNet's predictions were only logged.}
\label{fig:study-parts}
\end{figure*}

\section{Editing films that can change order}
\label{sec:authoring}

\subsection{Keep the opening and ending fixed}

A fixed opening and ending let us vary the middle while keeping the restaurant, the problem, and the outcome clear. Figure~\ref{fig:study-parts} shows how the edits fit into the study. We edited three television restaurant-inspection episodes into films. The opening introduces the restaurant and the inspection. The middle consists of short scenes that can be understood on their own, such as the discovery of a mouldy burger. We kept the original audio and cut each scene so that it could follow more than one other scene. The ending keeps the outcome of the inspection. We chose food and sanitation material and avoided injury and medical imagery. Other candidate material, including cleaning, restoration, hoarding and insect eating, had competing emotional content, pronounced camera motion, or too few scenes that could stand on their own.

Each film has six candidate scenes and plays four of them, beginning with the same first scene in all conditions. This leaves three choices that can change what plays, and two scenes that are never part of the fixed order. Table~\ref{tab:assets} gives each scene's subject, ratings, duration, role and reading interval.

\subsection{Use pilot ratings to order the choices}

In the pilot ratings, each fixed first scene had at least one gentler and one stronger scene available after it. Pilot viewers rated longer versions of the films, each about 7 min 46 s long and containing seven candidate scenes. The VR pilot had nine women and covered all six film orders. Films A and B each had nine VR ratings, and Film C had eight because one viewer watched that film on desktop. The desktop pilot had five women and four men for Films A and C, and four women and four men for Film B. Participants rated disgust from 0 to 5. They also reported other feelings and rated the context, transitions, ending, length, familiarity and technical problems. They marked when reactions occurred while reviewing the material, which helped us choose where to read the face in each scene.

The VR pilot's mean disgust ratings for the whole films were 4.00, 3.97 and 3.95 for A, B and C. The four scenes first selected for each film averaged 4.03, 3.95 and 3.94. Within each film, the gap between the highest and lowest scene mean was between 1.33 and 2.12 points. The films also elicited other feelings. The small VR pilot had only women, while the desktop groups were mixed, and this difference was one reason we ran the main study with women in VR.

Pilot ratings of the joins between scenes led us to shorten the scenes and revise how they were cut. Scenes in the deployed films lasted 8 to 28 s. We chose the mouldy burger (A02), expired chives (B03), and tomato sauce spill (C01) as the fixed first scenes. We used the VR pilot means, which came from the longer material, to order the available choices by disgust. These means set which scene counts as gentler or stronger, though scenes also differ in meaning and length.

\section{Reading the face and choosing a scene}
\label{sec:runtime}

FRAME VR reads the face, chooses a scene and plays the film on a Meta Quest Pro without a connected computer. The films are ordinary 2D video shown on a fixed virtual cinema screen. Viewers use controllers during onboarding and questionnaires. The application records the headset's numerical estimates of facial movements, or blendshape coefficients, with tracking validity and the time in the film and scene. The tracker supplied roughly 57 to 70 samples per second during the deployment. We also ran a pulse sensor and an external camera, but the analyses here use only the headset and questionnaire records.

\subsection{Read while the current scene is playing}

A range-of-motion calibration begins the session, and a 10 s recording of the viewer's resting face begins each film. When the marked interval for reading the face ends, the scene keeps playing while the application checks the samples recorded since the scene began. At least 80\% of those samples need valid tracking, timestamps must increase, and both the resting signal and the movement during the scene must vary enough to be usable. The expression model takes a 7.314 s window of 49 facial channels resampled to 512 evenly spaced samples. The window is centred on the peak of whole-face activity inside the marked interval and uses samples already recorded. Because the model needs this full window, the application did not attempt a reading when a scene's marked interval was shorter (A01 and C02).

For the scene choice, the rule takes the peak across four channels: the left and right estimates of the nose wrinkle and upper-lip raise. These movements are called facial action units 9 and 10 (AU9 and AU10). A peak below the resting mean plus one standard deviation is a weak reading, and a peak above the mean plus three standard deviations is a strong reading. Values in between, including the thresholds, form the middle band. An unusable reading is recorded with its reason. Each viewer is compared only with their own resting face, so the rule is not trained on other viewers, and every choice can be traced to one peak and two thresholds.

During each film, the headset also runs SpontNet, a compact model with 105,617 parameters trained on 5,602 annotated events from 45 headset wearers. It runs alongside playback and logs whether an expression occurred, whether it was brief or sustained, and its emotion label, while the rule uses only the lower-face peak to choose the scene. The model details and its separate recognition results are in Appendix~\ref{app:model}. Appendix~\ref{app:runtime} gives the input and playback specification.

\subsection{Choose among the unused scenes}

A strong reading asks for a gentler scene, so the rule takes the unused scene with the lowest pilot rating. A weak reading asks for a stronger scene, so the rule takes the unused scene with the highest rating. A reading in the middle band asks for a similar intensity, and the rule takes the closest rating. Each scene can play at most once. If there is no usable reading, the system uses the next unused scene in the fixed order, which we call the preset choice. This keeps the film playing, but it can raise or lower the intensity.

The system starts loading the selected scene on a second video surface while the current one ends. It records the first frame, the gap at the cut, and decoder and frame-stall notifications.

To check that SpontNet fits within a scene, we timed it. Across its 72 successful runs, including those in final scenes, the median time from preprocessing to model output was 486 ms, the 90th percentile was 1,066 ms, and the range was 249 to 1,898 ms. Preprocessing took a median of 327 ms, and the model itself took a median of 146 ms. All 72 outputs came from Adaptive, and 71 of them used captured facial samples alone. The remaining one also drew on samples from the resting recording.

\subsection{Record the request and what plays}

Each decision record keeps the current scene, the reading or the reason it could not be used, the direction requested, the unused scenes and their ratings, and the scene played. We subtract the current scene's pilot rating from the next scene's. Changes within 0.15 points count as similar. A match means the next scene's pilot rating was higher, lower or similar as requested. A match describes the cut in the film, while the viewer rates the feeling separately. This definition lets us tell a match from a preset choice or an unavailable direction using the same scene list. The final scene has a reading interval but no next scene to choose. The counts below come from our original study report, and Section~\ref{sec:scope} states what the records support.

\section{Studying the films with viewers}
\label{sec:study}

\subsection{Face-responsive, stored-replay and fixed-order films}

Each of the 36 viewers watched all three films, one in each condition. In \textbf{Static}, the scenes played in a fixed order that followed each original episode: A02, A01, A03, A04 for Film A, B03, B01, B04, B06 for Film B, and C01, C02, C04, C05 for Film C. In \textbf{Adaptive}, the current viewer's reaction could choose the next scene. In \textbf{Yoked}, the system replayed a reserved order for the same film without using the current viewer's reactions to choose scenes. This reserved order was the Adaptive order of an earlier viewer for the same film and study settings, or a preset order when no earlier order was available. The headset recorded estimates of facial movement in all conditions and used readings for selection only in Adaptive. Condition labels were hidden.

All viewers were women at one site. We excluded sessions in which talking, distraction or other behaviour compromised the facial recording. Among the viewers who remained, every film session entered the comparison, including films with failed readings or unavailable scenes. Most viewers (23 of 36) had never used a VR headset, six wore glasses in the headset, and 32 reported zero motion sickness at baseline. When asked how often they watch deliberately disgusting media, 3 said never, 21 rarely and 12 sometimes. Three of the 35 viewers who answered the familiarity question knew the source material.

We assigned films and conditions on a balanced schedule, but after exclusions the sessions were spread unevenly across films and presentation positions (Table~\ref{tab:balance}). The table counts films and positions separately. All viewers supplied the film and scene ratings and the four final film-comparison items. The later ranking, familiarity, personalisation and noticing items were available for 35 viewers because one session lacked those answers. For two viewers with conflicting saved answers, we used the corrected version.

\subsection{Ask how it felt and whether it suited the viewer}

We asked whether the level of disgust suited the experience a viewer wanted. Before the first film, the viewer read how to use the scale and answered two comprehension questions, each with one retry. After each film, the suitability question offered five choices, from much too mild to much too strong, with just right in the middle. For responses numbered 1 to 5, distance from just right is $|v-3|$. It is 0 for just right, 1 for a bit too mild or strong, and 2 for much too mild or strong. This score treats neighbouring answers as equally spaced. We also report all five categories and the number who chose exactly just right.

Viewers rated sickness, attention, ease of following, picture or sound problems, whether the film felt like separate pieces, overall disgust, headset discomfort, and whether any scene was hard to see. The attention and following questions were single items informed by work on narrative engagement~\cite{busselle2009measuring}. The sickness and technical questions adapted existing formats~\cite{keshavarz2011validating,itu2023p910}, and we kept headset discomfort separate from sickness~\cite{hirzle2021critical}. The viewer then saw thumbnails of the scenes in the order actually played, rated disgust from 0 to 10, and said whether each scene had too little, about the right amount, or too much disgust, or whether they were unsure. A final survey asked viewers to compare the three films and ended by asking whether any film seemed to change in response to them. Appendix~\ref{app:questionnaire} gives the wording and response options.

\subsection{Compare ratings from the same people}

We first counted the changes the system delivered, then compared ratings under the three conditions as implemented. Our main exploratory comparison used each viewer's distance from just right in Adaptive and Yoked. The 95\% intervals for mean differences used 100,000 resamples of participants. A second analysis asked whether a viewer chose exactly just right. This analysis adjusts for the film, when it played, and stable differences among viewers (participant fixed effects). Its interval grouped each person's answers and used a $t$ distribution with 35 degrees of freedom. A wider family of 36 tests, 12 outcomes times three condition pairs, used Holm's correction for multiple comparisons. Appendix~\ref{app:outcomes} lists all the recorded measures and available results.

\section{What the films delivered and what viewers reported}
\label{sec:results}

Every film finished on the headset, and the per-viewer rule delivered most of the requests it could act on. We report delivery first, then what viewers said about the films.

\subsection{The rule matched 36 of 108 choices overall, 36 of 51 with a usable reading}
\label{sec:delivery}

Of the 51 choices with a usable reading, 36 matched the request (71\%), which is 36 of the 108 Adaptive choices (33\%). The system fell back on the preset order in 57 choices because it had no usable reading. The other 15 had a usable reading but played a scene outside the requested direction. Our study report links these misses to the scenes left in the scene list. Most requests for a change asked for a gentler scene (37 of 45), and 8 asked for a stronger one. The 36 matches were 28 gentler, 4 stronger and 4 similar choices. The 57 preset choices made the next scene stronger 36 times, gentler 16 times and similar 5 times.

\begin{table}[t]
\centering
\small
\caption{Delivery from facial readings to scenes played. The bottom block gives the main result: 36 of 51 choices with a usable reading matched the request, or 36 of 108 overall. The 36 Adaptive films contributed 144 reading intervals, including the final interval in each film, and 108 next-scene choices. A match means the next scene's pilot rating moved in the requested direction. For a similar-intensity request, the change had to be within 0.15 points. With no usable reading, the next unused scene in fixed order played.}
\label{tab:delivery}
\begin{tabularx}{\linewidth}{@{}>{\raggedright\arraybackslash}Xr@{}}
\toprule
Observation & Count \\
\midrule
All conditions: film sessions completed & 108/108 \\
Stored-order replays that matched the reserved order & 36/36 \\
Adaptive films with a different order from Static & 23/36 \\
\midrule
Adaptive reading intervals accepted & 72/144 (50\%) \\
Adaptive intervals rejected or too short & 72/144 (50\%) \\
\midrule
Next-scene choices with a usable reading & 51/108 \\
Requested direction matched, of all choices & 36/108 (33\%) \\
Requested direction matched, of usable readings & 36/51 (71\%) \\
Preset choice because no reading was usable & 57/108 (53\%) \\
Reading usable, requested direction missed & 15/108 (14\%) \\
\bottomrule
\end{tabularx}
\end{table}

Half of the 144 Adaptive reading intervals yielded a usable result. Among the 72 that did not, 27 failed the resting-face checks, 21 had flat or near-zero signals, and 24 occurred where the edits left too little time. The 72 accepted readings were 54 strong, 11 weak and 7 in the middle band. These counts include readings in the last scene, so they show where readings were lost across the whole film, whereas the 57 preset choices count only the losses that changed what played next. Eleven of the 36 viewers had no usable reading in any of their four intervals, and 12 had no choice that matched the requested direction. The Adaptive order differed from Static in 23 of the 36 films, a count that includes order changes made by the preset rule.

Every film finished, giving 108 complete film sessions and 432 scene ratings. All 36 Yoked films played their reserved orders, of which 33 came from earlier Adaptive viewers and 3 were preset. Of the 33 orders from earlier viewers, 29 could be checked against their source sessions, and all 29 matched.

\subsection{Suitability after accounting for film and position}
\label{sec:suitability}

Neither suitability result shows a reliable Adaptive advantage. After adjusting for film and presentation position, the Adaptive minus Yoked difference in the share of viewers answering exactly just right was 23.8 percentage points, with a 95\% interval from $-2.8$ to $+50.3$, so this comparison was inconclusive. This estimate concerns the share answering exactly just right, not the distance from it.

\begin{figure}[t]
\centering
\includegraphics[width=\linewidth]{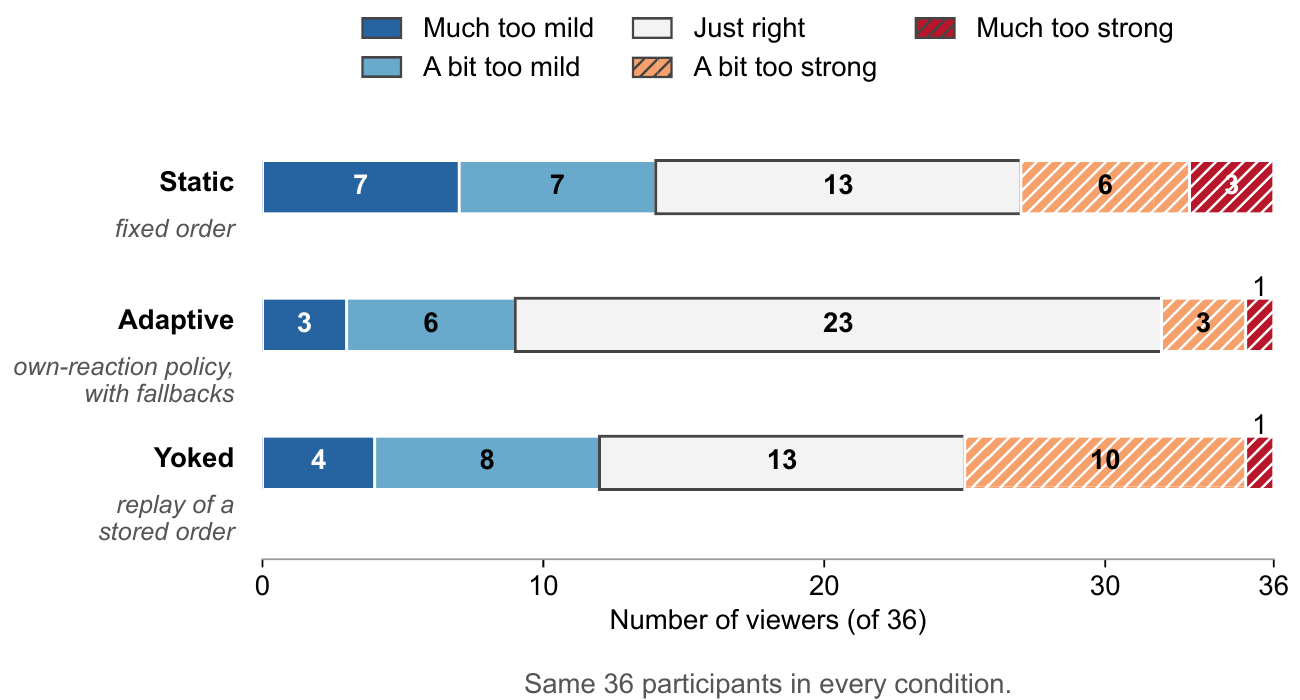}
\caption{All film-intensity suitability ratings from the same 36 viewers. Each bar runs from much too mild to much too strong. After adjusting for film and position, the Adaptive minus Yoked difference in exactly just right was 23.8 percentage points, with a 95\% interval from $-2.8$ to $+50.3$, which is inconclusive. The white segments show just right answers (Adaptive 23, Static 13, Yoked 13).}
\label{fig:suitability}
\end{figure}

In Adaptive, 23 of the 36 viewers chose just right, compared with 13 in each of the other two conditions (Figure~\ref{fig:suitability}). The unadjusted Adaptive minus Yoked difference was 27.8 percentage points.

The unadjusted mean distance from just right was 0.47 in Adaptive, 0.78 in Yoked, and 0.92 in Static. The paired Adaptive minus Yoked difference was $-0.31$ points, with a 95\% interval of $[-0.58,-0.03]$. Adaptive was closer for 16 viewers, Yoked for 7, and 13 tied. Adaptive minus Static was $-0.44$ with an interval of $[-0.69,-0.19]$, and Yoked minus Static was $-0.14$ with an interval of $[-0.42,+0.14]$. These distance intervals do not adjust for film or position. None of the wider set of 36 outcome comparisons met the significance threshold after Holm correction.

Adaptive received fewer too-mild ratings even though 37 of the 45 requests for a change asked for a gentler scene. It also drew fewer too-strong ratings. Gentler requests should, if anything, raise too-mild ratings. The preset order, which made 36 choices stronger, may explain part of the drop. The ratings reflect both kinds of choice and the film each viewer saw.

\subsection{Ratings of following, attention and the joins}

Average ease of following ranged from 5.97 to 6.22 out of 7. Ratings of whether the film felt like separate pieces ranged from 3.94 to 4.39 on the same scale. Attention ranged from 5.42 to 5.83. Table~\ref{tab:film_outcomes} gives the means and standard deviations for all three conditions. Across conditions, 31 of the 108 film sessions had at least one stall notification, an application record of a possible playback problem. Mean sickness after the films ranged from 1.14 to 1.47 on a 0 to 20 scale. Mean headset discomfort ranged from 1.39 to 1.50 on a 0 to 5 scale. Nearly half the viewers (17 of 36) missed at least one suitability comprehension question even after retries, and they remain in the comparisons. This limits how far the suitability results can be read.

\begin{table*}[t]
\centering
\caption{Film ratings from the same 36 people in Static (fixed order), Adaptive (the current viewer's reactions and preset choices when needed), and Yoked (stored-order replay). Cells give mean (standard deviation across viewers). Higher scores mean more of the named quantity. For distance from just right, 0 means just right and 2 means much too mild or strong.}
\label{tab:film_outcomes}
\begin{tabular*}{\textwidth}{@{}l@{\extracolsep{\fill}}rrrr@{}}
\toprule
Measure & Range & Static & Adaptive & Yoked \\
\midrule
Distance from just right & 0--2 & 0.92 (0.81) & 0.47 (0.70) & 0.78 (0.68) \\
Felt disgust & 0--10 & 3.81 (2.90) & 4.00 (2.66) & 3.72 (2.66) \\
Held attention & 1--7 & 5.42 (1.18) & 5.83 (0.77) & 5.44 (0.97) \\
Easy to follow & 1--7 & 5.97 (1.08) & 6.14 (0.76) & 6.22 (0.59) \\
Felt like separate pieces & 1--7 & 4.39 (1.96) & 3.94 (1.82) & 3.97 (1.75) \\
\bottomrule
\end{tabular*}
\end{table*}

\subsection{Few viewers thought a film had responded}
\label{sec:noticing}

The rule changed what played, yet only 4 of the 35 respondents thought a film changed in response to them. Another 25 were unsure and 6 said no. On a 0 to 10 scale, the average sense that a film responded personally was 2.54, and 17 of the 35 respondents gave it zero. These answers show whether viewers suspected that any film responded, not whether they picked out the Adaptive one. Only the decision records show what played. A film can make the requested cut while most viewers do not report noticing a response. Whether a response should be noticeable is therefore a design decision for responsive media, and one that needs testing.

\subsection{Strong facial readings did not always mean strong disgust}
\label{sec:movement}

Of the 54 strong readings, 21 came from scenes rated at most 2 out of 10 for disgust, and 7 from scenes the viewer said had too little disgust. Across 72 accepted readings from 25 viewers, larger peaks went with more reported disgust on average (pooled Pearson $r=0.34$, 95\% interval $[0.07,0.54]$ from 20,000 resamples grouped by viewer, and Spearman $\rho=0.24$). The 72 scenes without a usable reading averaged 4.0 out of 10 for disgust, compared with 3.8 for scenes with a reading, so missing readings were not limited to milder scenes.

Across all conditions, viewers judged the intensity about right in 181 of the 432 scene ratings, too little in 132, too much in 92, and were unsure in 27. Mean disgust on the 0 to 10 scale was 1.58 over the 132 too-little ratings, 4.16 over the 181 about-right ratings, and 8.16 over the 92 too-much ratings. The rule assumes that a strong facial reading calls for relief, so that assumption needs testing against what viewers want.

\section{What the film material could deliver}
\label{sec:check}

A filmmaker can run the scene check on any responsive film's scene list before any viewer sees the film. It shows how one responsive cut can use up the time, or the gentler scenes, that the next cut needs. Each of our films offered a gentler choice after its fixed first scene, but taking the gentlest one caused trouble at the next cut. In Film A, the rule follows the mouldy burger with the fly strip, which leaves too little time for the headset to read the next reaction, and Film C has the same problem. Film B's gentlest scene leaves a long enough interval to attempt a reading, but no unused scene is rated lower.

\subsection{Check the current choice and what it leaves}

After the study, we wrote a scene check, a program that reads the list of scenes, the pilot ratings, the fixed order, and the marked intervals for reading the face. For a given scene and a set of unused scenes, it first asks whether a scene in the requested direction exists, then follows the selection rule and checks the chosen scene's rating. If there is another choice after it, the program checks whether the selected scene leaves a marked interval of at least 7.314 s and whether an unused scene can still satisfy another request in the same direction. It also reports when the current interval is too short even to attempt a reading. The last choice only needs to supply the requested scene. The check needs only the scene list and the selection rule, so it runs before any viewing. We ran it on the three deployed films.

\begin{table}[t]
\centering
\small
\caption{Gentler choices after each film's fixed first scene. The second column gives the number of gentler unused scenes, then how many of those allow at least 7.314 s for a reading. The deployed rule chooses the lowest-rated unused scene. Even in Film B, where a long enough interval remains, there is no scene rated lower than the one chosen.}
\label{tab:anchor}
\begin{tabular}{@{}lccrl@{}}
\toprule
Film & Gentler / long enough & Chosen & Interval (s) & \shortstack[l]{What limits\\the next choice} \\
\midrule
A & 2 / 1 & A01 & 7 & Too short to read \\
B & 1 / 1 & B01 & 9 & No gentler scene \\
C & 1 / 0 & C02 & 6.5 & Too short to read \\
\bottomrule
\end{tabular}
\end{table}

\begin{figure}[t]
\centering
\includegraphics[width=\linewidth]{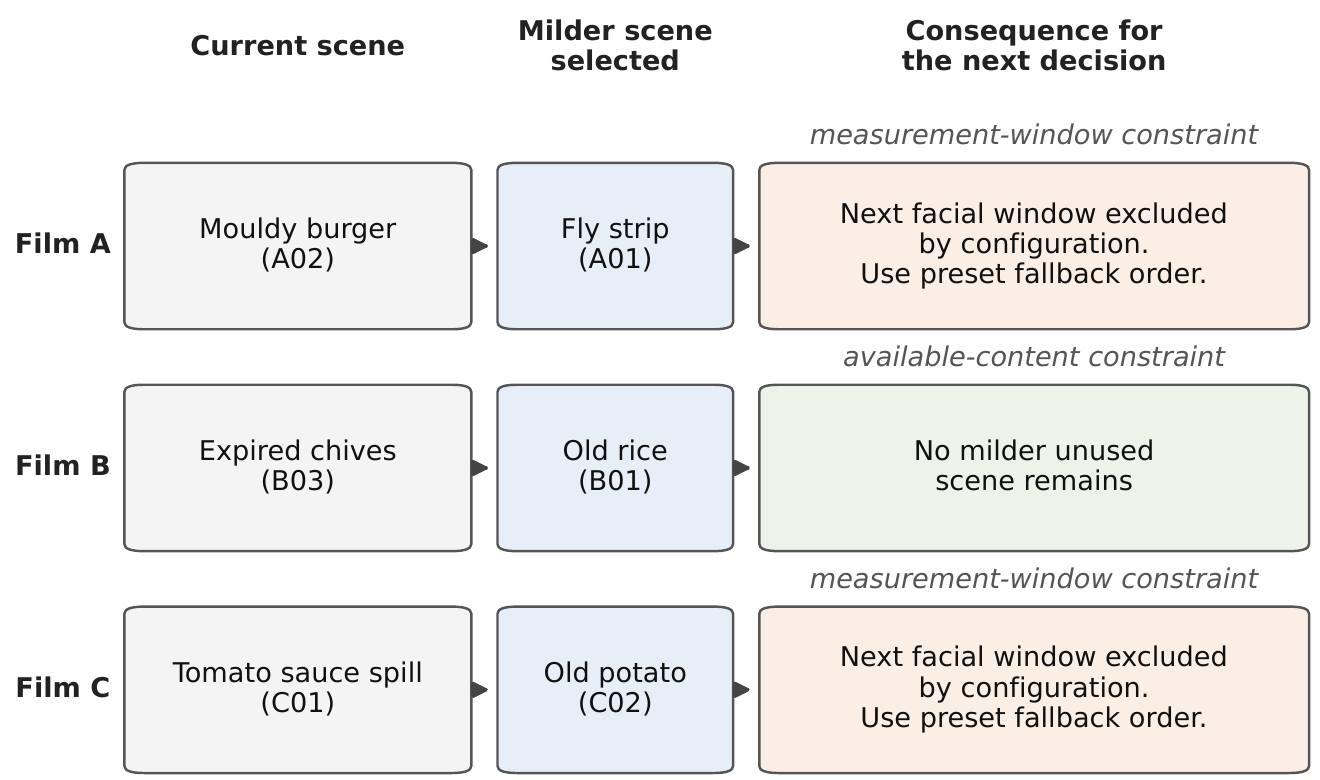}
\caption{What follows when a strong reading after each film's fixed first scene makes the rule choose the gentlest unused scene. Read each row left to right. In Films A and C, the chosen scene leaves too little time to read the face, so the next choice follows the preset order. In Film B, the chosen scene leaves enough time but uses up the only gentler scene. In the artwork, milder means gentler (a lower mean pilot rating), and facial window means reading interval. Scene names are shortened from Table~\ref{tab:assets}.}
\label{fig:content}
\end{figure}

In Film A, the size of the step matters (Table~\ref{tab:anchor} and Figure~\ref{fig:content}). Its mouldy burger scene (A02) has a pilot rating of 4.22. The rule chooses the fly strip scene (A01), rated 3.11, which leaves just 7 s for the next reading. A06, flies from bagged salad, is rated 3.22. It also moves the film towards gentler content and leaves 8 s for a reading. The step from A06 to A01, however, is only 0.11 points, within the deployed 0.15-point band for similar intensity. If the rule chose A06, A01 would then count as similar rather than gentler, and no gentler step would remain. Film B has only one gentler scene after its fixed first scene. Film C also has only one, and that scene's reading interval lasts 6.5 s. In Films B and C, the scenes available at the first choice limit what the system can do.

\subsection{Follow all the possible orders}

Picking the lowest or highest rating uses up the extreme in the requested direction, whichever scene the film is currently playing. If the rule chooses the lowest unused rating, no unused scene can be rated lower than the new scene. The same applies in the stronger direction when it chooses the highest. So responding to a second request in the same direction may need a different selection rule as well as enough scenes with distinct ratings.

We extended the scene check to all 26 combinations of current scene and unused scenes in each film. Each film starts with its fixed first scene, followed by five possible second scenes and 20 possible ways to reach a third scene before the last cut. We considered each of the four inputs at each step: a weak reading (stronger scene), a strong reading (gentler scene), a reading in the middle band (similar scene), and no usable reading. We kept every tied choice and compared the deployed definition of similar intensity, within 0.15 points, with exact ordering, in which only equal ratings count as similar. The selection rule stays the same in both calculations.

Each film has six points where a later choice still follows the current one. Under the deployed definition, a smaller step would meet the current request, leave a long enough interval and preserve another option in the same direction at 0 of these 6 points in Film A, 2 of 6 in B and 3 of 6 in C. The same counts hold for gentler and for stronger requests. Under exact ordering, the counts were 4 of 6, 3 of 6 and 3 of 6 for gentler requests, and 3 of 6 in each film for stronger requests. For some cases, the two definitions label the same selected cut with a different direction. The tables in Appendix~\ref{app:structural} give all cases, including those with no suitable alternative.

\section{Lessons for making a responsive film}
\label{sec:lessons}

\paragraph{Give the story a stable opening and ending.}
Viewers generally found these films easy to follow, but their ratings also suggest that they could feel the joins. Let the opening introduce the situation and the ending resolve it. Cut each scene in between so that a viewer can follow its references and understand why it is there after more than one preceding scene. Review the candidate joins for changes in speaker, sound or context that a single-scene review would miss.

\paragraph{Leave time to read the viewer after each cut.}
A scene can be right for the current request and still prevent the next reading. A01 and C02 were gentler choices, but their marked intervals were too short for the next reading. Mark the reading interval on the final edit, then check how much of the scene remains for choosing and loading the next one. Decide what should play if there is no reading, and check how that choice changes the film. In this deployment, the preset order could make the film stronger.

\paragraph{Save a scene for the next request.}
Check what remains after the choice, especially when someone may want to move further in the same direction. Film A had a gentler alternative with room to read (A06), but the step from A06 to A01 counts as similar under the deployed 0.15-point band, so it would keep a gentler step in reserve only with a narrower band. Films B and C had only one gentler choice at the start. If the available scenes include a suitable smaller step, taking it can preserve a larger step for later. If the list has no smaller step, change the edit or add material. Check the joins for these paths as well as their ratings and reading intervals.

\paragraph{Check what played and ask what the viewer understood.}
The rule delivered 36 of the 51 requests that had a usable reading, or 36 of 108 choices overall, yet few viewers thought a film had responded to them. Keep the request, the scenes that were available, and the scene that played together in each decision record. For preset choices, keep the reason the reading was unusable. Report the matches among all choices first, then the matches among usable readings. If a comparison uses a stored order, check its replay against the reserved order and record where the order came from. Decide what viewers should be told about the sensing and the possibility of a response, and ask afterwards what they understood.

\paragraph{Ask what viewers want as well as what they feel.}
In these films, viewers said that a scene had too little disgust more often than too much, and some strong facial readings happened in scenes rated low in disgust. A strong reaction can be part of why someone watches. Ask what the viewer felt and whether that intensity fit the experience they wanted. To test a rule that asks for relief, also ask which direction they would want the film to take next.

\section{Scope and future work}
\label{sec:scope}

This study shows that a film system on a standalone headset can respond to a viewer's face, and where that response can break down. It covers 36 women at one site, three films cut from television, and one headset with a fixed virtual screen. The pilot groups were small. The suitability comparison is exploratory and applies to the films as implemented, including the preset choices. Uneven assignment after exclusions, single custom questions, and viewers' difficulty with the suitability scale limit it. Resampling or grouping by viewer accounts for repeated answers from the same viewer, but Yoked orders can also link viewers to one another, and those analyses do not account for that dependence. Each viewer contributes several decisions and scene ratings, so these are not independent. The study has no established power or equivalence claim.

The participant results are aggregates from our original study report. The records do not identify the final list of included sessions or the files that mark corrected answers. We therefore did not reanalyse the numbers, since that would have required substituting an earlier cohort or choosing which answer versions to use. For the same reason, this paper does not report:
\begin{itemize}
\item how many sessions were excluded and why
\item the cause of each missed request and how failures were spread across viewers
\item how corrected answers or missed comprehension questions affect the results
\item each viewer's film-by-position assignment and how many viewers answered each item
\item the questionnaire version for each session and which earlier viewer supplied each Yoked order
\item the full set of comparisons and the exact specification of the adjusted model
\end{itemize}
 Four of the earlier sessions behind Yoked orders were not available for the replay comparison. The drop in too-mild ratings in Adaptive cannot be attributed to facial or preset choices. Timing beyond the model's successful runs, such as the time from a reading to the next scene and playback at the cut, is left for future work. Appendix~\ref{app:runtime} describes the scene-selection path from our study report.

These limits suggest several next steps. A new study with other viewers, headsets and genres would help show which lessons carry over. Purpose-written films could offer several compatible continuations at a cut. Ratings of the final edits and tests of their joins could show whether the pilot ordering and the sense of story survive the edit. A study with filmmakers could test whether running the scene check during editing helps them find and fix these problems. The scene check uses scene ratings and the length of each marked interval. A long enough interval can still have poor tracking, and a scene that satisfies the requested direction can still produce a poor join. The check counts possible cuts without weighting them by how often viewers would meet them, and it has not yet been tested as an aid to filmmakers, as an improved selection policy, or as a predictor of viewer benefit. The exact-ordering analysis treats the pilot means as exact, so it shows one scenario and gives no uncertainty. A device study could measure the time from reading to the first frame through repeated cuts, including failures. It could also test whether an application that uses only the lower-face channels can make the same decisions with shorter reading intervals.

The rule measures lower-face movement. A still face can accompany a strong feeling, and a small movement against a nearly flat resting signal can look large to this rule. The pooled association with disgust leaves open how well the measurement follows changes within a person. A further study could ask for the desired direction before testing the same rule, measure how expressive range and fit affect the reading, and compare a rule that keeps within a viewer's stated preferences. The expression benchmark used instructions to suppress visible reactions, one trained annotator to combine labels, and a deployed model from one training fold. Independent labelling and recordings during ordinary viewing would test how well those predictions carry over.

\section{Ethics and data availability}
\label{sec:ethics}

Both studies received institutional ethics approval and all participants provided informed consent. Participants could stop through a control in the headset or by speaking to the researcher. The application processed the facial coefficients on the headset. Even without the inward-facing camera images, facial movement can carry identifying information. FRAME VR data are not released. The figures use diagrams and schematic illustrations of the film screen.

\section{Conclusion}
\label{sec:conclusion}

FRAME VR shows that a face-responsive film system can run on a standalone headset. With 36 viewers, its per-viewer rule played a scene in the requested direction on 36 of 51 choices with a usable reading, and on 36 of 108 choices overall. SpontNet ran alongside playback on the same headset and logged its expression predictions. Only 4 of 35 respondents thought any film had responded to them, so viewers may not notice a requested cut, and what played and what viewers noticed need separate checks. The scene check shows, before anyone watches, where the rule can run out of a suitable change. In Film A, the gentlest scene after the mouldy burger left too little time to read the next reaction, so one requested cut left the film unable to respond at the next. For responsive films, the scene list, the time allowed to read a reaction, and the choices left after a cut are all part of what the filmmaker decides. Our five lessons turn these findings into steps that filmmakers can take.

\section*{Acknowledgements}
This work was supported by the Dolby Laboratories University Research Collaboration program. We thank all study participants.

\section*{AI use}
Anthropic coding agents were used to implement and debug parts of the training and evaluation pipelines and assist with text editing. The authors designed the study, verified all results and citations, and take full responsibility for this work.

\bibliographystyle{framevr-doi}
\bibliography{refs}
\onecolumn
\appendix

\section{Film scene list and worked choices}
\label{app:assets}

Table~\ref{tab:assets} lists all 18 scenes that could play. The fixed orders were A02, A01, A03, A04 for A, B03, B01, B04, B06 for B, and C01, C02, C04, C05 for C. A scene in the fixed order always remains if the film has a choice left. The VR pilot supplied nine ratings each for A and B and eight for C. The desktop pilot supplied five women's ratings for A and C, four for B, and four men's ratings for each film.

\begin{table}[htbp]
\centering\small
\caption{Every candidate scene, with mean disgust ratings from women in VR (V), women on desktop (W), and men on desktop (M). Ratings are from 0 to 5 and refer to the longer pilot material. V uses nine ratings for A and B and eight for C. W uses five for A and C and four for B. M uses four per film. Duration and the start and end of the interval for reading the face are in seconds for the final edit. The role shows whether the scene always played first, belonged to the rest of the fixed order, or was available only as an extra scene. Within each film, rows are sorted by VR rating, lowest first.}\label{tab:assets}
\begin{tabular}{@{}lp{0.29\linewidth}rrrrll@{}}
\toprule
Scene & Content & Duration (s) & V & W & M & Interval (s) & Role \\
\midrule
A01 & Fly strip & 13 & 3.11 & 3.40 & 1.50 & 0--7 & Fixed \\
A06 & Flies from bagged salad & 22 & 3.22 & 4.40 & 3.25 & 11--19 & Extra \\
A02 & Mouldy burger & 13 & 4.22 & 4.40 & 2.25 & 2--11 & First \\
A04 & Cockroaches & 26 & 4.33 & 4.60 & 2.50 & 11--24 & Fixed \\
A03 & Rat evidence & 28 & 4.44 & 4.20 & 2.50 & 16--24 & Fixed \\
A07 & Tomato intended for service & 25 & 4.44 & 4.60 & 4.00 & 11--20 & Extra \\
B01 & Old pre-cooked rice & 19 & 3.11 & 2.50 & 1.50 & 2--11 & Fixed \\
B03 & Expired chives & 28 & 3.56 & 3.25 & 3.50 & 8--20 & First \\
B07 & Old refried beans & 27 & 4.22 & 4.25 & 5.00 & 6--14 & Extra \\
B04 & Old fish in liquid & 17 & 4.33 & 4.25 & 4.25 & 0--11 & Fixed \\
B06 & Green, slimy stored meat & 26 & 4.78 & 4.75 & 4.50 & 6--20 & Fixed \\
B05 & Oxidised stored meat & 25 & 4.78 & 4.75 & 3.75 & 8--16 & Extra \\
C02 & Old twice-baked potato & 8 & 2.62 & 3.00 & 3.00 & 0--6.5 & Fixed \\
C01 & Tomato sauce spill & 26 & 3.25 & 2.40 & 3.00 & 11--19 & First \\
C03 & Grease under equipment & 25 & 4.00 & 4.20 & 3.50 & 1--20 & Extra \\
C04 & Mouldy pesto & 21 & 4.38 & 3.80 & 4.00 & 3--19 & Fixed \\
C07 & Part-cooked, reheated chicken & 25 & 4.38 & 4.20 & 4.50 & 7--15 & Extra \\
C05 & Mould-spotted food remnant & 24 & 4.75 & 4.20 & 4.00 & 4--14 & Fixed \\
\bottomrule
\end{tabular}
\end{table}

\subsection{A checklist for editing and reviewing joins}
Give the first and last scenes enough context to establish and resolve the story. Keep each middle scene understandable if the viewer has only seen the opening and a different middle scene. Record references a viewer would need to recognise, changes in speakers and sound, and any dependency on an earlier scene. For every scene, keep its pilot rating alongside the rated version, the duration of the final edit, and the start and end of the interval when the headset reads the face. Check the joins the selection rule can produce, including those made when a reading is unusable. The pilot informed the edits and did not supply a rating for each final join.

\subsection{Trace a choice from request to screen}
Table~\ref{tab:storyboard} shows how the same film can make a gentler choice with a reading and a stronger one without a reading. From the opening of Film A's selectable middle, the available scenes are A01, A03, A04, A06 and A07. The lower rating of A06 and its 8 s interval make it an alternative to A01 after A02. The difference between A06 and A01 is only 0.11 points, so A01 counts as similar after A06 within the deployed 0.15-point band. It would count as gentler under exact ordering.

\begin{table}[!htbp]
\centering\small
\caption{Worked scene choices using the edited film material. Each row gives a possible choice with an assumed input and shows the change in pilot disgust rating. These examples show that a change in the played order can come from a reading or from the preset order.}
\label{tab:storyboard}
\begin{tabularx}{\linewidth}{@{}p{0.16\linewidth}p{0.16\linewidth}p{0.24\linewidth}>{\raggedright\arraybackslash}X@{}}
\toprule
Current scene & Assumed input & Selected scene and direction & What remains \\
\midrule
A02, first scene & Strong reading & A01, $3.11-4.22=-1.11$, gentler match & The 7 s interval is too short to read. No unused scene is rated lower. \\
A02, first scene & Unusable reading & A01 in fixed order, $-1.11$, gentler & The same scene plays through the preset choice. \\
A01, after A02 & Unusable reading & A03 in fixed order, $4.44-3.11=+1.33$, stronger & The preset choice raises the pilot-rated intensity. \\
B03, first scene & Strong reading & B01, $3.11-3.56=-0.45$, gentler match & The 9 s interval is long enough to attempt a reading. No unused scene is rated lower. \\
\bottomrule
\end{tabularx}
\end{table}

\section{Headset input, selection and playback records}
\label{app:runtime}

\subsection{Check the recording before choosing}

The scene rule uses the usable facial interval and the viewer's resting-face reference. The 7.314 s input and the 512 time samples are requirements of the expression model. The film application did not attempt a reading in scenes whose marked interval was shorter than this length (A01 and C02). These scenes could still be chosen and played. Tracking coverage, variation in the resting recording, and variation in the lower-face channels determine whether the signal itself is usable. The path described in our study report is captured samples, signal checks, an eligible interval, the AU peak and response band, scene selection, then loading and playback. SpontNet takes a separate input branch and logs predictions. Its expression probability has no threshold in the scene-selection rule.

Available historical code waits for model output and includes expression gating and an abort path. The selection rule in this paper is the one described in our study report, and Section~\ref{sec:scope} notes what the code we still have can confirm.

The resting recording lasts 10 s per film. The application checks for increasing timestamps and at least 80\% valid tracking. A flat reading has a peak of zero or a window standard deviation below $10^{-6}$. A resting recording is unusable if its standard deviation is below $10^{-4}$. The rule takes the peak across the left and right nose wrinkle and upper-lip raise during the marked interval. With resting mean $\mu$ and standard deviation $\sigma$, a peak below $\mu+\sigma$ asks for stronger content and a peak above $\mu+3\sigma$ asks for gentler content. A peak between or on those boundaries asks for similar intensity.

\subsection{Prepare the model input}

We subtract the neutral median, divide by the viewer's 95th-percentile calibration range, and mask channels whose range is below 0.005. The input uses samples already recorded centred on a whole-face movement peak inside the marked interval. Interpolation gives 512 evenly spaced values per channel. One execution included padding taken from the resting recording. Table~\ref{tab:channels} shows the available channel map. The floating-point input has 512 time samples of 49 facial channels and a final zero-filled slot. Resampling sets a common time grid but does not recover movements missed during recording.

\begin{table}[htbp]
\centering\small
\caption{Channel order in the available input map. Read the names left to right, one row at a time. L and R mean left and right. B and T mean bottom and top. NoseWrinkler and UpperLipRaiser estimate the selection rule's lower-face movements. The map comes from the available source code.}\label{tab:channels}
\begin{tabular}{@{}lll@{}}
\toprule
Channel & Channel & Channel \\
\midrule
EyesClosedL & EyesClosedR & UpperLidRaiserL \\
UpperLidRaiserR & LidTightenerL & LidTightenerR \\
InnerBrowRaiserL & InnerBrowRaiserR & OuterBrowRaiserL \\
OuterBrowRaiserR & BrowLowererL & BrowLowererR \\
CheekRaiserL & CheekRaiserR & CheekPuffL \\
CheekPuffR & NoseWrinklerL & NoseWrinklerR \\
UpperLipRaiserL & UpperLipRaiserR & LipCornerPullerL \\
LipCornerPullerR & DimplerL & DimplerR \\
LipCornerDepressorL & LipCornerDepressorR & LipTightenerL \\
LipTightenerR & LipPressorL & LipPressorR \\
LipStretcherL & LipStretcherR & ChinRaiserB \\
ChinRaiserT & LipFunnelerLB & LipFunnelerRB \\
LipFunnelerLT & LipFunnelerRT & LipPuckerL \\
LipPuckerR & MouthLeft & MouthRight \\
JawDrop & JawSidewaysLeft & JawSidewaysRight \\
JawThrust & LipsToward & LowerLipDepressorL \\
LowerLipDepressorR &  &  \\
\bottomrule
\end{tabular}
\end{table}

\subsection{Select, load and record what played}

The pseudocode gives the choices described for the playback system and how the scene check handles ties.

\begin{quote}
\fontencoding{OT1}\fontfamily{cmtt}\selectfont\small
If four scenes have played, record the final reading and stop choosing.\par
U := the unused scenes for this film.\par
If the reading is unusable, choose the first scene from the fixed order in U.\par
Else if the reading is weak, choose an unused scene with the highest rating.\par
Else if the reading is strong, choose an unused scene with the lowest rating.\par
Otherwise choose an unused scene with the closest rating to the current scene.\par
For the scene check, keep every choice tied on the selection rule.\par
Difference := the selected scene's rating minus the current scene's rating.\par
The selected scene is stronger above 0.15 and gentler below -0.15.\par
Between those values, including the boundaries, it is similar.\par
Compare the direction played with the request whenever there was a usable reading.\par
Mark the selected scene as used and start loading it while the current scene ends.
\end{quote}

The scene check keeps every choice tied on the selection rule, so it covers whichever tie-breaking rule playback used. Table~\ref{tab:schema} gives the fields used to follow a reading and choice through the film and link it to the viewer's answers.

\begin{table}[!htbp]
\centering\small
\caption{Records used to describe a scene choice. The facial interval identifies a reading opportunity, and the decision identifies a choice that can change the film. Playback records identify what played, and questionnaire records link the viewer's answers to that film or scene. Fields can differ by record version.}
\label{tab:schema}
\begin{tabularx}{\linewidth}{@{}p{0.21\linewidth}>{\raggedright\arraybackslash}X@{}}
\toprule
Record & Fields \\
\midrule
Facial interval & Film, scene, played position, and whether this is the final scene. \\
Reading & Tracking and signal checks, resting-face summary, lower-face peak, response band or reason for an unusable reading. \\
Decision & Requested direction, unused scenes and pilot ratings, available directions, scene selected, and change in rating. \\
Model timing & Start and end of processing and model execution, and execution status where recorded. \\
Playback & First frame, time to prepare it, gap from the previous scene, interval between frames, and stall notifications. \\
Questionnaire & Item, film and condition, scene where applicable, and revision of the saved answer. \\
Stored-order replay & Source type (earlier viewer or preset), source session when applicable, and the order reserved for replay. \\
\bottomrule
\end{tabularx}
\end{table}

\subsection{What the records can establish}

Our check of the source found a model file that matched the checksum in the surviving metadata. The deployment numbers in this paper are those of our study report.

The records have fields for reading attempts, preparation, first frames and gaps. Field names and coverage vary by record version. Missing or sentinel timing values do not mean instantaneous execution. A timing comparison needs an explicit common monotonic clock. We do not treat notifications as independently observed visible freezes.

\section{Questionnaire items and response options}
\label{app:questionnaire}

The wording below is from the available version that matches our study. Viewers answered with the controllers and could revise answers before submitting. They saw the scenes in the order played, with each scene's position and a thumbnail. The screen showed neither the condition names nor the scene intensity labels. The questions about whether any film responded came after all films.

\subsection{Before the first film}
\paragraph{Resting sickness} Right now, how much motion sickness or nausea do you feel?

Integer control 0--20. No sickness at all to Severe sickness.

\paragraph{Viewing habits} How often do you watch films or shows that you would describe as gross or stomach-turning?

Responses: Never; Rarely; Sometimes; Often; Very often.

\paragraph{Seeking intensity} I enjoy films that push me further than feels comfortable.

Responses: 1 - Strongly disagree; 2 - Disagree; 3 - Somewhat disagree; 4 - Neither agree nor disagree; 5 - Somewhat agree; 6 - Agree; 7 - Strongly agree.

\paragraph{Prior headset use} Have you used a VR headset before today?

Responses: No; Yes.

\paragraph{Vision correction} Are you wearing vision correction inside the headset?

Responses: No; Glasses; Contact lenses; Prefer not to say.

\paragraph{Suitability explanation} Understanding the answers

``After each film we will ask whether the disgusting moments were too mild, just right, or too strong for the experience you wanted. 'Just right' means the amount suited you. It does not mean the film was very disgusting, and it does not mean you liked the film.''

Responses: Continue.

\paragraph{Suitability check 1} If the disgusting moments were exactly the amount you wanted, which answer would you choose?

Responses: Much too mild; A bit too mild; Just right; A bit too strong; Much too strong. Correct response: Just right. A wrong answer triggers the explanation and one retry. It does not block continuation.

\paragraph{Suitability check 2} If the film was far more disgusting than you wanted, which answer would you choose?

Responses: Much too mild; A bit too mild; Just right; A bit too strong; Much too strong. Correct response: Much too strong. A wrong answer triggers the explanation and one retry. It does not block continuation.

\subsection{After each film}
\paragraph{Motion sickness} Right now, how much motion sickness or nausea do you feel?

Integer control 0--20. No sickness at all to Severe sickness.

\paragraph{Intensity fit} For the experience you wanted, the disgusting moments in this film were...

Responses: Much too mild; A bit too mild; Just right; A bit too strong; Much too strong.

\paragraph{Attention} This film held my attention.

Responses: 1 - Strongly disagree; 2 - Disagree; 3 - Somewhat disagree; 4 - Neither agree nor disagree; 5 - Somewhat agree; 6 - Agree; 7 - Strongly agree.

\paragraph{Ease of following} It was easy to follow what was happening in this film.

Responses: 1 - Strongly disagree; 2 - Disagree; 3 - Somewhat disagree; 4 - Neither agree nor disagree; 5 - Somewhat agree; 6 - Agree; 7 - Strongly agree.

\paragraph{Picture and sound} Did you notice any problems with the picture or sound, such as freezing, stuttering or glitches?

Responses: No problem; Noticeable but not annoying; Slightly annoying; Annoying; Very annoying.

\paragraph{Fragmentation} At some point, this felt like separate pieces joined together rather than one complete film.

Responses: 1 - Strongly disagree; 2 - Disagree; 3 - Somewhat disagree; 4 - Neither agree nor disagree; 5 - Somewhat agree; 6 - Agree; 7 - Strongly agree.

\paragraph{Overall disgust} Overall, how disgusting was this film?

Integer control 0--10.

\paragraph{Headset discomfort} How uncomfortable did the headset feel on your head or face during this film?

Responses: None; Very slight; Slight; Moderate; Strong; Extreme.

\paragraph{Clarity} Were there any moments that you could not see clearly?

Responses: No; Yes.

\paragraph{Unclear scenes} Which moments could you not see clearly?

Options are the scenes that played. This question follows a Yes to the clarity question. Multiple selections permitted. Not sure is exclusive.

\subsection{For each played scene, in viewing order}
\paragraph{Scene disgust} How disgusted or grossed out did this moment make you feel?

Integer control 0--10.

\paragraph{Scene fit} For the experience you wanted, the level of disgust in this moment was...

Responses: Too little; About right; Too much; Not sure.

\subsection{After all films, before debrief}
\paragraph{Best flow} In which film did the moments flow best from one to the next?

Responses: 1st film; 2nd film; 3rd film; No difference; Not sure.

\paragraph{Best intensity fit} In which film was the level of disgust closest to what you wanted?

Responses: 1st film; 2nd film; 3rd film; No difference; Not sure.

\paragraph{Most attention} Which film held your attention the most?

Responses: 1st film; 2nd film; 3rd film; No difference; Not sure.

\paragraph{Watch again} Which film would you most like to watch in this format again?

Responses: 1st film; 2nd film; 3rd film; None; Not sure.

\paragraph{Disgust ranking} Rank the films from most to least disgusting.

Responses: 1st film  ->  2nd film  ->  3rd film; 1st film  ->  3rd film  ->  2nd film; 2nd film  ->  1st film  ->  3rd film; 2nd film  ->  3rd film  ->  1st film; 3rd film  ->  1st film  ->  2nd film; 3rd film  ->  2nd film  ->  1st film.

\paragraph{Familiarity} Before today, had you seen any of these episodes or most of these scenes?

Responses: No; Not sure; Yes.

\paragraph{Perceived response} How much did it feel like any film was responding to you personally?

Integer control 0--10.

\paragraph{Whether a film responded} Did you think any film changed based on your reactions?

Responses: No; Not sure; Yes.

\paragraph{Which film responded} Which film or films did you think changed based on your reactions?

Responses: 1st film; 2nd film; 3rd film; Not sure. This question follows Yes or Not sure to whether a film responded. Multiple selections permitted. Not sure is exclusive.

\subsection{Scoring and availability}

The suitability response runs from 1 to 5. The distance score is $|v-3|$, and the separate just-right count includes only responses of 3. For attention, ease of following and fragmentation we report each agreement item on its original 1 to 7 scale. Headset discomfort is coded from 0 to 5 in the order above. The disgust guide labels 0 as not at all, 5 as moderately and 10 as extremely. Sickness uses 0 to 20.

The reporting covers 36 viewers on the film and scene ratings and the first four comparison questions, and 35 on the later ranking, familiarity, personalisation and noticing questions. The available conditional follow-up counts are listed in Appendix~\ref{app:outcomes}.

\section{All outcomes and assignments}
\label{app:outcomes}

Table~\ref{tab:outcome_inventory} includes favourable, unfavourable and unavailable results. The analysis comprised 36 comparisons. The film and presentation-position counts and the suitability estimates follow the table. Figure~\ref{fig:suitability} gives the full rating distribution.

\small
\begin{longtable}{@{}>{\raggedright\arraybackslash}p{.22\linewidth}>{\raggedright\arraybackslash}p{.29\linewidth}>{\raggedright\arraybackslash}p{.43\linewidth}@{}}
\caption{All recorded study measures with any available sample counts and results. S, A and Y denote Static (fixed order), Adaptive (the current viewer's reactions and preset choices as needed), and Yoked (stored-order replay). Each person watched all three conditions.}\label{tab:outcome_inventory}\\
\toprule
Measure & Scale and availability & Result \\
\midrule
\endfirsthead
\toprule Measure & Scale and availability & Result \\
\midrule\endhead
\bottomrule\endfoot
Film intensity fit & five ordered categories; distance score and choice of exactly just right; 36 per condition & distance means S/A/Y 0.92/0.47/0.78; counts from much too mild to much too strong: S 7/7/13/6/3, A 3/6/23/3/1, Y 4/8/13/10/1 \\ 
Film attention & 1 to 7 agreement on one question; 36 per condition & means S/A/Y 5.42/5.83/5.44; SD 1.18/0.77/0.97 \\ 
Film following & 1 to 7 agreement on one question; 36 per condition & means S/A/Y 5.97/6.14/6.22; SD 1.08/0.76/0.59 \\ 
Film fragmentation & 1 to 7 agreement on one question; 36 per condition & means S/A/Y 4.39/3.94/3.97; SD 1.96/1.82/1.75 \\ 
Film disgust & felt disgust, 0-10; 36 per condition & means S/A/Y 3.81/4.00/3.72; SD 2.90/2.66/2.66 \\ 
Motion sickness & sickness, 0-20 adapted anchors; pre-film baseline separate; item count unavailable & post-film means 1.14 to 1.47; baseline zero for 32 \\ 
Headset discomfort & six categories, 0-5; item count unavailable & condition means 1.39 to 1.50 \\ 
Technical problems & five-category picture/sound impairment; item count unavailable & condition statistics unavailable \\ 
Scene clarity and which scenes & yes/no and conditional scene thumbnails; item count unavailable & statistics unavailable \\ 
Scene disgust & felt disgust, 0-10; 432 & accepted-window mean 3.8 vs no-read mean 4.0; Section~\ref{sec:movement} gives the association for usable readings \\ 
Scene fit & too little/about right/too much/not sure; 432 & 132/181/92/27; category disgust means 1.58/4.16/8.16 for first three \\ 
Best flow & best flow by viewing order/no difference/not sure; 36 & condition counts unavailable \\ 
Best intensity fit & closest desired intensity by viewing order/no difference/not sure; 36 & condition counts unavailable \\ 
Most attention & most attention by viewing order/no difference/not sure; 36 & condition counts unavailable \\ 
Watch again & watch again by viewing order/none/not sure; 36 & condition counts unavailable \\ 
Disgust ranking & six permutations of viewing order; 35 & condition statistics unavailable \\ 
Familiarity & no/not sure/yes; 35 & 3 familiar \\ 
Perceived response & 0-10; 35 & mean 2.54, 17 zeros \\ 
Whether a film responded & no/not sure/yes; 35 & 6/25/4 \\ 
Which film responded & conditional multiple choice; item count unavailable & correct-film identification unavailable \\ 
Suitability comprehension & all attempts plus retries; 36 & 17 failed at least 1 check after retries; results by comprehension status unavailable \\ 
Viewing habits, seeking intensity, headset use, vision correction & baseline context; item counts unavailable & Section~\ref{sec:study} gives the available counts \\ 
\end{longtable}
\normalsize

\begin{table}[!ht]
\centering\small
\caption{Film and position assignments. The top three rows show how often each condition came first, second or third. The bottom three rows show how often it was paired with Film A, B or C. Each row sums to 36 viewings. Static is fixed order, Adaptive uses the current viewer's reaction or the preset order, and Yoked is stored-order replay.}\label{tab:balance}
\begin{tabular}{@{}lrrr@{}}
\toprule
Condition & Position 1 / Film A & Position 2 / Film B & Position 3 / Film C \\
\midrule
Static: positions & 15 & 9 & 12 \\
Adaptive: positions & 12 & 12 & 12 \\
Yoked: positions & 9 & 15 & 12 \\
Static: films & 12 & 11 & 13 \\
Adaptive: films & 14 & 11 & 11 \\
Yoked: films & 10 & 14 & 12 \\
\bottomrule
\end{tabular}
\end{table}

\begin{table}[!ht]
\centering\small
\caption{Suitability comparisons between playback conditions. Negative distances mean the first condition was closer to just right. Distance intervals come from resampling viewers without adjustment for film or position. The first row adjusts the comparison of exactly just right for film and position. Its interval crosses zero. Each comparison uses the same 36 viewers.}\label{tab:contrasts}
\begin{tabular}{@{}llrl@{}}\toprule
Outcome & Contrast & Estimate & 95\% interval \\\midrule
Exactly just right, adjusted (percentage points) & Adaptive minus Yoked & $23.8$ & $[-2.8,+50.3]$ \\
Exactly just right, unadjusted (percentage points) & Adaptive minus Yoked & $27.8$ & Unavailable \\
Distance, 0--2 & Adaptive minus Yoked & $-0.31$ & $[-0.58,-0.03]$ \\
Distance, 0--2 & Adaptive minus Static & $-0.44$ & $[-0.69,-0.19]$ \\
Distance, 0--2 & Yoked minus Static & $-0.14$ & $[-0.42,+0.14]$ \\
\bottomrule
\end{tabular}
\end{table}

\section{Full scene check}
\label{app:structural}

Each film has 26 distinct combinations of current scene and unused scenes at which a choice can be assessed: one at the fixed first scene, five at the next scene, and 20 at the scene before the final choice. For the first six states with a later choice, the selected scene must meet three conditions to preserve a continuation: match the current request, allow 7.314 s for the next reading, and leave an unused scene that can satisfy another request in the same direction. The last 20 states need only supply the requested scene. States with short current intervals are present in the full counts.

Tables~\ref{tab:audit_deployed_015} and~\ref{tab:audit_exact_order} separate the scenes available from the scenes selected. Availability counts states with an unused scene that fits the request. The match counts include all tied selections. The continuation column counts earlier states in which all, some, or none of those selections meet the three conditions. The final column splits the earlier states where none meet them into those with an unselected scene that meets them and those with no such alternative.

Following the selection rule from the fixed first scene, while allowing an unusable reading at any point and a facial request only after a long enough interval, reaches 13 of the 26 states in A, 12 of 26 in B, and 7 of 26 in C under either definition. These are possible film states under the interval rule, not observed paths or verified executable states. Tracking, decoder readiness, cut timing and joins are outside this abstraction. The tables cover all 26 states per film.

\begin{table}[!ht]
\centering\small
\caption{Scene check using the deployed 0.15-point band for similar intensity. Availability counts states with a fitting unused scene out of all 26 states, including ones with short intervals. Match fractions count selected scenes that fit the request out of all selections, including ties. The continuation column covers the six states that still have a later choice. For each, it records whether all, some or none of the tied selections meet three conditions: they match the current request, leave 7.314 s for another reading, and leave another scene in the requested direction. The last column splits states with no continuation into those with a suitable unselected alternative and those with none. An unusable reading carries no request.}\label{tab:audit_deployed_015}
\begin{tabular}{@{}llrrrr@{}}
\toprule
Film & Input or request & Available & Matches / rows & Continue: all / some / none & Alternative / none \\
\midrule
A & Gentler & 16 & 16/26 & 0/0/6 & 0/6 \\
A & Stronger & 11 & 18/36 & 0/0/6 & 0/6 \\
A & Similar & 24 & 27/29 & 4/0/2 & 0/2 \\
A & No usable reading & n/a & n/a (26 rows) & n/a & n/a \\
B & Gentler & 19 & 19/26 & 0/0/6 & 2/4 \\
B & Stronger & 16 & 26/36 & 0/0/6 & 2/4 \\
B & Similar & 16 & 16/28 & 0/0/6 & 0/6 \\
B & No usable reading & n/a & n/a (26 rows) & n/a & n/a \\
C & Gentler & 20 & 20/28 & 0/0/6 & 3/3 \\
C & Stronger & 19 & 21/31 & 0/0/6 & 3/3 \\
C & Similar & 8 & 8/33 & 0/0/6 & 0/6 \\
C & No usable reading & n/a & n/a (26 rows) & n/a & n/a \\
\bottomrule
\end{tabular}
\end{table}
\begin{table}[!ht]
\centering\small
\caption{Scene check using exact ordering, in which only equal ratings count as similar. Columns are as in Table~\ref{tab:audit_deployed_015}. An unusable reading carries no request.}\label{tab:audit_exact_order}
\begin{tabular}{@{}llrrrr@{}}
\toprule
Film & Input or request & Available & Matches / rows & Continue: all / some / none & Alternative / none \\
\midrule
A & Gentler & 20 & 20/26 & 0/0/6 & 4/2 \\
A & Stronger & 16 & 26/36 & 0/0/6 & 3/3 \\
A & Similar & 8 & 8/29 & 0/0/6 & 0/6 \\
A & No usable reading & n/a & n/a (26 rows) & n/a & n/a \\
B & Gentler & 20 & 20/26 & 0/0/6 & 3/3 \\
B & Stronger & 16 & 26/36 & 0/0/6 & 3/3 \\
B & Similar & 8 & 8/28 & 0/0/6 & 0/6 \\
B & No usable reading & n/a & n/a (26 rows) & n/a & n/a \\
C & Gentler & 20 & 20/28 & 0/0/6 & 3/3 \\
C & Stronger & 19 & 21/31 & 0/0/6 & 3/3 \\
C & Similar & 8 & 8/33 & 0/0/6 & 0/6 \\
C & No usable reading & n/a & n/a (26 rows) & n/a & n/a \\
\bottomrule
\end{tabular}
\end{table}

Selecting the lowest or highest unused rating never leaves a further move in that direction under either definition. For similar-intensity requests, the deployed rule preserves a continuation in 4 of the 6 earlier Film A states and in 0 of 6 in B and C. Exact ordering yields 0 of 6 in each film. In either case, no unselected scene restores a missed similar-intensity continuation.

The tied ratings are A03 and A07 at 4.44, B05 and B06 at 4.78, and C04 and C07 at 4.38. The check also keeps scenes that are equally close to the current rating. Across the four inputs, including all ties, some selected cuts change their direction label when exact ordering replaces the 0.15-point band. The tables retain both definitions rather than treating close means as precise intensity steps.

\begin{table}[!htbp]
\centering\small
\caption{Choices when there is no usable reading. A row counts scene choices made by the fixed order at the 26 possible states for each film under one definition of similar intensity. These are counts from the film material. There is no face-based request to match.}
\label{tab:audit_unusable}
\begin{tabular}{@{}llrrr@{}}
\toprule
Definition & Film & Gentler & Similar & Stronger \\
\midrule
Within 0.15 points & A & 13 & 7 & 6 \\
Within 0.15 points & B & 19 & 1 & 6 \\
Within 0.15 points & C & 18 & 1 & 7 \\
Equal ratings only & A & 18 & 1 & 7 \\
Equal ratings only & B & 19 & 0 & 7 \\
Equal ratings only & C & 18 & 1 & 7 \\
\bottomrule
\end{tabular}
\end{table}

\section{Facial-expression recordings and model}
\label{app:model}

SpontNet is a compact model that runs on the headset alongside playback, and it was strongest at detecting whether an expression occurred. Its emotion predictions were much weaker, and a classifier that knew only the stimulus timeline did better at naming the emotion. This appendix describes the separate recording study, the model, and its evaluation. The film rule in Section~\ref{sec:runtime} uses the viewer's nose and upper-lip movements.

\subsection{Record movements and let viewers label their replay}

The Spont-VR recording study followed the BlendME protocol~\cite{sribabu2026blendme}. The Quest Pro recorded 70 coefficients, 63 facial and 7 eye, at a nominal 45 samples per second. Participants completed an 11-step range-of-motion calibration. We subtracted the neutral median of each channel, divided by the wearer's 95th-percentile range, and masked channels with a range below 0.005. The input to the model kept 49 facial channels and excluded the seven eye coefficients.

A separate group of 45 adults watched 14 clips from an emotion-elicitation library~\cite{schaefer2010assessing}. The pilot assessed 37 candidate clips in three batches with 15 raters each. The selected set had two clips for each of six emotion categories, with one extra each for anger and surprise. We asked viewers to watch as if in public and to avoid showing what they felt~\cite{gross1993suppression}. After each clip they reported the emotion and rated its intensity on a 5-point scale.

The participants later watched an avatar replay of their face and labelled movements, following the approach used by BlendME and related retrospective methods~\cite{sribabu2026blendme,potts2025retrosketch}. A detector proposed moments when a facial channel changed. A trained annotator combined the viewer's in-headset report and replay labels, the pilot stimulus labels, the reaction moments marked by three independent annotators, and the detector's proposed interval. For each event the annotations mark its start, peak and end. The 5,602 events comprised 2,858 brief expressions, 2,456 longer expressions, and 288 movements labelled as non-expressions. The median duration was about 330 ms for brief expressions and 1.2 s for longer ones. The 5,314 expression events carried labels for sadness, happiness, disgust, fear, anger or surprise. Table~\ref{tab:datasets} places these labels and recording conditions alongside earlier collections.

\begin{table}[!htbp]
\centering\small
\caption{Expression datasets by recorded signal, task and labels. Headset coefficients are numerical estimates of facial movement. RGB-D includes colour and depth video. Suppression means viewers were asked to avoid displaying feelings. Brief and longer expressions follow the definitions in each dataset.}
\label{tab:datasets}
\begin{tabularx}{\linewidth}{@{}l>{\raggedright\arraybackslash}p{0.16\linewidth}>{\raggedright\arraybackslash}X>{\raggedright\arraybackslash}X@{}}
\toprule
Dataset & Signal & Recording setting & Labels \\
\midrule
CASME II~\cite{yan2014casme} & Facial video & Films with suppression & Brief expressions \\
SAMM~\cite{davison2018samm} & Facial video & Films with suppression & Brief expressions \\
CAS(ME)$^3$~\cite{li2023cas} & RGB-D facial video & Films and mock-crime tasks & Brief and longer expressions \\
DFME~\cite{zhao2024dfme} & Facial video & Films with suppression & Brief expressions \\
EmoHeVRDB~\cite{ortmann2024emojiherovr} & Quest Pro coefficients & Expressions posed to emoji & Emotion \\
Natural vs. acted~\cite{rahman2026generalized} & Quest Pro coefficients & Videos and acted expressions & Emotion \\
BlendME~\cite{sribabu2026blendme} & Quest Pro coefficients & Films with suppression & Brief expressions labelled from replay \\
Spont-VR & Quest Pro coefficients & Films with suppression & Brief, longer and non-expression events. Emotion for expression events \\
\bottomrule
\end{tabularx}
\end{table}

\subsection{Describe the movements across facial regions}

SpontNet learns time patterns in each part of the face, then combines the parts. Its input is a 7.314 s window of facial movements resampled by linear interpolation to 512 evenly spaced samples, or about 70 samples per second. This sets a common time grid for irregular headset samples. The extra input slot contains zeros. Stimulus identity and timing are excluded from the main model tests.

\begin{figure}[!htbp]
\centering
\includegraphics[width=\linewidth]{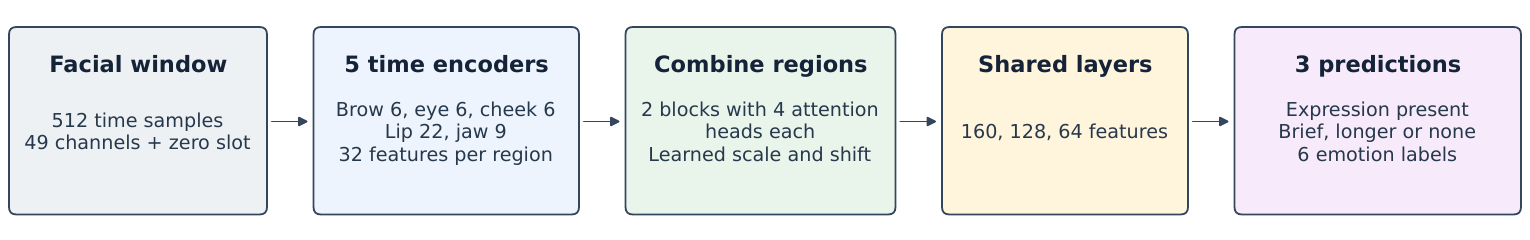}
\caption{SpontNet turns the headset's numerical estimates of movements into three predictions. The encoders learn changes over time in five facial regions. The attention blocks combine the region summaries, learned scale and shift terms adjust them, and the shared layers feed separate predictions of expression presence, duration class, and emotion. The model has 105,617 trainable parameters and its predictions are logged alongside the rule that chooses scenes.}
\label{fig:model}
\end{figure}

The brow, eye area, cheek, lip and jaw each have a separate convolutional encoder. A convolution here learns how patterns change over nearby points in time. Each encoder produces a 32-number summary. Attention combines these summaries across facial regions in two blocks with four heads per block. Feature-wise linear modulation (FiLM) scales and shifts the features to account for stable differences among wearers during training. A shared network of fully connected layers feeds three prediction branches (Figure~\ref{fig:model}). The model has 105,617 trainable parameters. Training used focal cross-entropy to give harder examples more weight, participant weighting, label smoothing, and mixed-precision arithmetic.

The architecture study considered temporal convolution, transformer, and bidirectional state-space models ranging from 26,000 to 3.5 million parameters. We used a 14-wearer subset to examine the regional grouping and attention. The full benchmark below holds each wearer out in turn and uses three training seeds. The headset ran one model fold.

\subsection{SpontNet recognises movement more reliably than emotion}

\begin{table}[!htbp]
\centering\small
\caption{SpontNet and comparison models on held-out wearers. Expression presence distinguishes expression from non-expression. Expression type distinguishes brief, longer and non-expression events. Macro-F1 ranges from 0 to 1. Emotion accuracy is the percentage correct among six labels. The table uses the benchmark scores from holding each of 45 wearers out in turn, with three training seeds. Emotion labels cover 5,314 expression events. SpontNet's 95\% intervals from resampling the folds were 0.457 to 0.524 for type macro-F1 and 28.3 to 36.2\% for emotion accuracy. Fold-level prediction denominators, aggregation weights and an interval for expression presence are not available.}
\label{tab:benchmark}
\begin{tabular}{@{}lrrr@{}}
\toprule
Predictor & Presence macro-F1 & Type macro-F1 & Emotion accuracy \\
\midrule
SpontNet & 0.685 & 0.491 & 32.2\% \\
Gradient boosting & n/a & 0.499 & 31.5\% \\
Random forest & n/a & 0.453 & 31.8\% \\
Multilayer perceptron & n/a & 0.459 & 26.2\% \\
Logistic regression & n/a & 0.447 & 29.7\% \\
Support vector machine & n/a & 0.420 & 30.1\% \\
Most common class & 0.480 & 0.230 & About 24\% \\
Uniform random & 0.400 & 0.290 & 16.7\% \\
\bottomrule
\end{tabular}
\end{table}

Expression presence had a macro-F1 of 0.685, compared with 0.480 for always predicting the most common class and 0.400 for uniform random prediction (Table~\ref{tab:benchmark}). Macro-F1 weights each class equally. The expression labels account for 5,314 of the 5,602 annotated events, so accuracy alone would give an incomplete picture. SpontNet's three-way expression-type macro-F1 was 0.491, close to the 0.499 from gradient boosting on summaries of the facial recordings. Emotion accuracy was 32.2\% and emotion macro-F1 was 0.146.

The timeline alone gave emotion accuracy of 70.3\%, and the face-plus-timeline model reached 43.3\%. A classifier can therefore predict labels from the viewing schedule without reading a face. The main benchmark gives the model facial estimates only. In a separate test on 1,421 posed sequences from EmoHeVRDB~\cite{ortmann2024emojiherovr}, the combined predictions of the face-only models from all folds reached 32.4\% accuracy and 0.210 macro-F1 without fine-tuning.

\section{Original study diagrams}
\label{app:original-diagrams}

These diagrams keep the artwork from the original study report. They show the project stages and the supporting expression model. The scene-selection rule and deployment evidence are described in the main text.

\begin{figure}[!ht]
\centering
\includegraphics[width=0.78\linewidth]{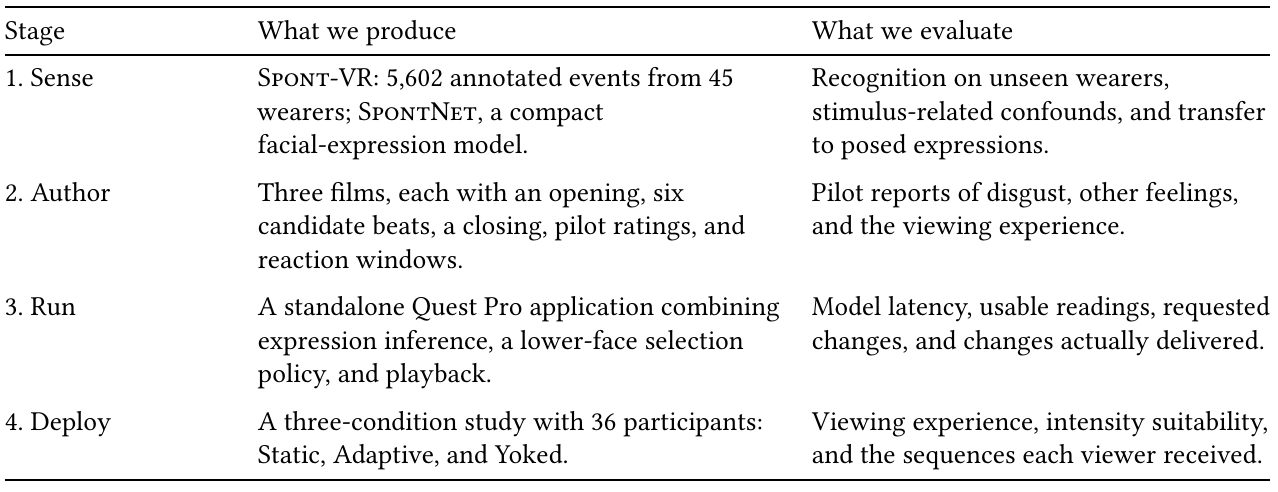}
\caption{Original study overview. Read across each row for the output and its evaluation. The expression study used 45 wearers and the separate film deployment used 36 viewers. Each film has six candidate scenes and plays four. The trained model logs predictions, and the per-viewer lower-face rule selects scenes. The artwork is reproduced from the original study-overview diagram. In the artwork, beats are scenes, reaction windows are reading intervals, and the lower-face selection policy is the per-viewer rule.}
\label{fig:original-overview}
\end{figure}

\begin{figure}[!ht]
\centering
\includegraphics[width=0.78\linewidth]{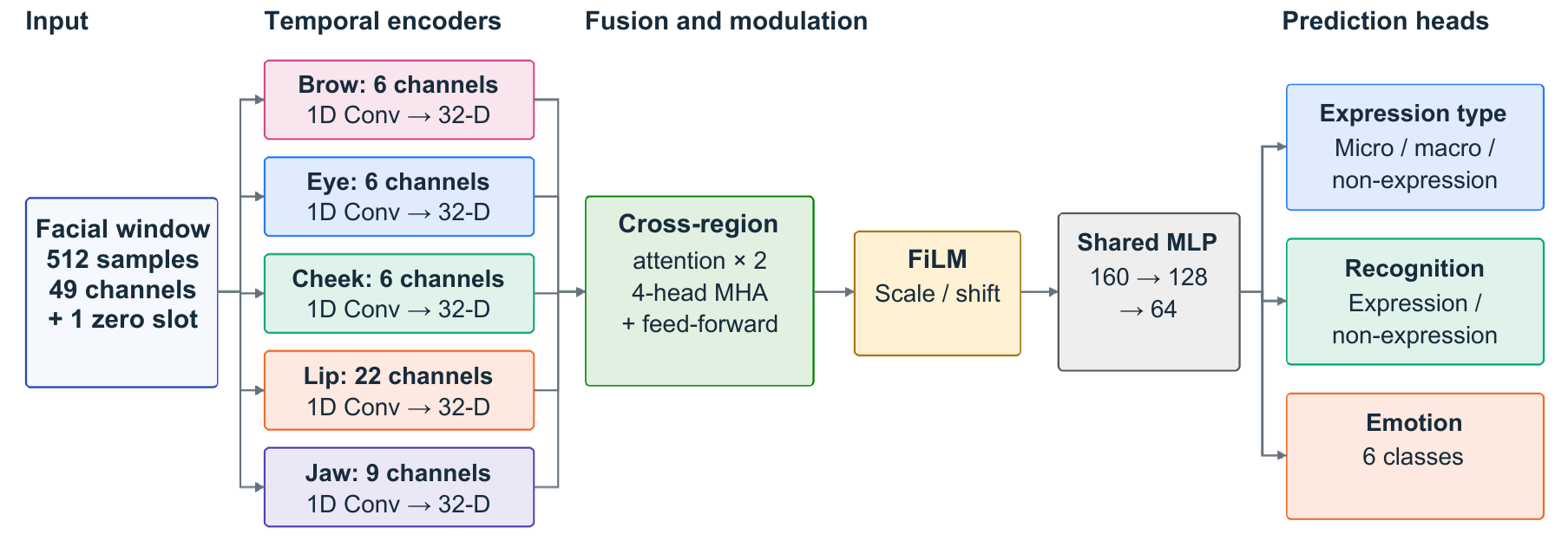}
\caption{Original SpontNet architecture. The input contains 512 time samples of 49 facial channels plus a zero-filled slot. The model has 105,617 trainable parameters. Its outputs are logged separately from scene selection. The artwork is reproduced from the original model diagram. Micro and macro in the artwork mean brief and longer expressions, and Recognition is expression presence.}
\label{fig:original-model}
\end{figure}

\end{document}